\documentclass[11pt]{article}

\usepackage{graphicx}
\usepackage{epstopdf}
\usepackage{hyperref}
\usepackage{booktabs}
\usepackage{natbib} 
\usepackage{amssymb}
 \usepackage{amsthm}
 \usepackage{amsmath}
 \usepackage{geometry}

\title{Asymmetric connectedness of stocks: How does bad and good volatility spill over the U.S. stock market?\thanks{An earlier version of this paper was circulated and cited under the title ``Asymmetric volatility spillovers: Revisiting the Diebold-Yilmaz (2009) spillover index with realized semivariance". We are indebted for many useful comments, suggestions, and discussions to Ionut Florescu, seminar participants at Modeling High Frequency Data in Finance 5 (New York, 2013) and Computational and Financial Econometrics (London, 2013). Support from the Czech Science Foundation under the 14-24129S grant is gratefully acknowledged. The research leading to these results has received funding from the European Union's Seventh Framework Programme (FP7/2007-2013) under grant agreement No. FP7-SSH- 612955 (FinMaP).}}

\author{Jozef Barun\'{\i}k\thanks{Institute of Economic Studies, Charles University, Opletalova 21, 110 00, Prague, Czech Republic and the Institute of Information Theory and Automation, Academy of Sciences of the Czech Republic, Pod Vodarenskou Vezi 4, 182 00, Prague, Czech Republic. Phone: +420 776 259 273. Email: barunik@utia.cas.cz.} \and Ev\v{z}en Ko\v{c}enda\thanks{Corresponding author: CERGE-EI, Charles University and the Czech Academy of Sciences, Politickych veznu 7, 11121 Prague, Czech Republic; CESifo, Munich; IOS, Regensburg; The William Davidson Institute at the University of Michigan Business School; CEPR, London; and the Euro Area Business Cycle Network. Email: evzen.kocenda@cerge-ei.cz.} \and Luk\'a\v{s} V\'acha\thanks{Institute of Economic Studies, Charles University, Opletalova 21, 110 00, Prague, Czech Republic and Institute of Information Theory and Automation, Academy of Sciences of the Czech Republic, Pod Vodarenskou Vezi 4, 182 00, Prague, Czech Republic. Email: vachal@utia.cas.cz.}} 

\date{}

\begin{document}
\maketitle

\begin{abstract}
\noindent Asymmetries in volatility spillovers are highly relevant to risk valuation and portfolio diversification strategies in financial markets. Yet, the large literature studying information transmission mechanisms ignores the fact that bad and good volatility may spill over at different magnitudes. This paper fills this gap with two contributions. One, we suggest how to quantify asymmetries in volatility spillovers due to bad and good volatility. Two, using high frequency data covering most liquid U.S. stocks in seven sectors, we provide ample evidence of the asymmetric connectedness of stocks. We universally reject the hypothesis of symmetric connectedness at the disaggregate level but in contrast, we document the symmetric transmission of information in an aggregated portfolio. We show that bad and good volatility is transmitted at different magnitudes in different sectors, and the asymmetries sizably change over time. While negative spillovers are often of substantial magnitudes, they do not strictly dominate positive spillovers. We find that the overall intra-market connectedness of U.S. stocks increased substantially with the increased uncertainty of stock market participants during the financial crisis.\\

\end{abstract}

\section{Introduction and related literature}

The presence of asymmetric volatility in financial markets has long been recognized in the literature \citep{black1976,christie1982stochastic,pindyck1984,french1987expected}. On the other hand, asymmetries in volatility spillovers have not yet received the same attention, despite the fact that the proper quantification of such asymmetries is highly relevant to risk valuation and portfolio diversification strategies \citep{GarciaTsafack2011}. One reason is the lack of a methodology that is versatile enough. In this paper we contribute to the literature in two ways: (i) we suggest how to quantify asymmetries in volatility spillovers due to bad and good volatility\footnote{\cite{bartram2012us} discuss the concepts of bad and good volatility. Good volatility of stocks results from conditions that enable firms to be more productive, while bad volatility might originate, for example, from high political risk that causes economic destabilization and decline.} and (ii) empirically show their substantial differences among the US stocks.

One of the stylized facts associated with financial markets reveals that the interdependence of markets exhibits asymmetries as large negative returns are more correlated than large positive returns \citep{longinsolnik2001,AngChen2002}. When contemporaneous returns and their conditional volatility exhibit negative correlation, then a stronger reaction to negative news results in asymmetric volatility of the assets \citep{wu2001determinants}. The causal link often leads to volatility spillovers, which tend to increase the idiosyncratic risk that diminishes gains from portfolio diversification \citep{Kanas2001}. In addition, \cite{Amonlirdviman2010} explicitly show that the asymmetry in the correlations of returns decreases the gains from portfolio diversification.

Volatility spillovers across markets are larger when market interdependence is high. At the same time, market returns tend to be more correlated when volatility increases and key periods of high volatility are associated with market downturns or crashes \citep{wu2001determinants}. Market volatility, especially in association with crisis development, then spills quickly across markets \citep{diebold2012better}. 
Asymmetry in volatility on financial markets implies that past returns are negatively correlated with present volatility \citep{bekaert2000asymmetric}. Since volatility is transferred across markets via spillovers, it is worth assuming that volatility spillovers exhibit asymmetries as well. Both volatility and its spillovers represent informative measures relevant for risk valuation and portfolio diversification strategies \citep{GarciaTsafack2011}. Therefore, asymmetric effects have to be quantified properly since it matters a lot whether volatility is originally of the bad or good type \citep{bartram2012us}, as differences in bad and good volatility exhibit distinctively different impacts on asset prices \citep{Segal2014}.

Our paper is underlined by several strands of the literature on volatility and volatility spillovers. The volatility of financial markets has become an intense topic of research since the emergence of the conditional heteroskedasticity models of \cite{engle} and \cite{bollerslev1987}. Subsequently, it has been recognized that volatility propagates in an asymmetric manner: this feature has been formalized in an exponential GARCH model in \cite{nelson1991conditional} and later formulated in a leverage effect ARCH model in \cite{glosten1993relation} as well as in a threshold ARCH model in \cite{zakoian1994threshold}. The use of these models led to an expansive body of empirical evidence confirming the asymmetric effect of negative versus positive returns on volatility. Specifically, in a number of early empirical studies it was shown that the volatility of stock markets tends to increase following negative or positive news but reacts more sensitively to bad news (see for example \cite{koutmos1995asymmetric, braun1995good}). More precisely, in the literature researchers usually identify volatility increases after negative returns but the effect of positive or close-to-zero returns is less clear \citep{ederington2010asymmetric}.\footnote{In order to explain the asymmetries well-known in the stock market, \cite{black1976} and \cite{christie1982stochastic} introduced the leverage effect as a source of asymmetric volatility in stocks. The leverage effect hypothesis is based on the idea that negative returns, due to the decline in the stock price, raises financial leverage, which increases the stock riskiness and volatility. Further, \citep{pindyck1984} and \cite{french1987expected} proposed that the asymmetry in volatility might also reflect the existence of time-varying risk premia: when volatility is priced, its expected increase leads to the growth of the required return on equity, which is followed by a decrease in stock price. The mechanism has been labelled as the volatility feedback effect. Both leverage and feedback effects are related to volatility as a risk proxy and may work at the same time. \cite{bekaert2000asymmetric} provide a summary of selected empirical studies on asymmetric volatility that explain the asymmetry via the above effects. \cite{wu2001determinants} developed a model that nested the two effects and empirically showed the leverage effect to be an important source of asymmetric volatility and that the volatility feedback effect has a stronger effect than was shown before.}

Later on, research on volatility expanded from a univariate to a multivariate framework, beginning with the bivariate GARCH model proposed by \cite{engle1995multivariate}. In the next step, \cite{engle2001theoretical} and \cite{engle2002dynamic} devised a Dynamic Conditional Correlation (DCC) GARCH model representing a non-linear combination of univariate GARCH models. To account for the asymmetry in a multivariate context, \cite{cappiello2006asymmetric} introduced the asymmetric DCC (ADCC) specification to account for asymmetries in the conditional variances and in the conditional correlations.

Research on volatility on financial markets became increasingly connected with the issue of how the volatility in one asset propagates to the volatility of other asset(s): volatility spillovers. 
Similar to volatility, much of the recent research on volatility spillovers employs versions of the GARCH model (for example \cite{beirne2013volatility,li2013modelling,lin2013examining}, among others). However, the ability to measure spillovers by those types of models is limited, namely in their lack of spillover dynamics. Recent developments in the literature related to spillovers introduced a new way to capture volatility spillovers more effectively.

In their seminal work, \cite{diebold2009measuring} developed a volatility spillover index (the DY index) based on forecast error variance decompositions from vector autoregressions (VARs) to measure the extent of volatility transfer among markets.\footnote{The DY index has been rapidly adopted in the relevant literature, for example, by \cite{mcmillan2010return,yilmaz2010return,bubak2011volatility,fujiwara2012asian,kumar2013returns}.} The methodology has been further improved in \cite{diebold2012better}, who used a generalized VAR framework in which forecast-error variance decompositions are invariant to variable ordering. The improved DY index is able to measure both total and directional volatility spillovers. \cite{diebold2012better} applied their new methodology to several classes of U.S. financial assets and showed that limited cross-market volatility spillovers before the 2007 crisis increasingly build up after the global financial crisis fully unfolded. Details on the DY index are provided later in section \ref{sec:metodology}.

Despite its versatility, the DY index does not distinguish the potential asymmetry in spillovers that originate due to bad and good volatility. However, new techniques recently improved our ability to estimate and explain volatility, including asymmetric effects. In response to the sharply rising use of high-frequency and ultra-high-frequency data, \cite{andersen2001distribution} and \cite{barndorff2002econometric} initially proposed to estimate quadratic variation as the sum of the squared returns and developed an estimator of the realized variance (RV). This development then enabled the emergence of yet a new measure of volatility--realized semivariance--due to \cite{shephard2010measuring}. Realized semivariance ($RS$) measures the variation of the change in the asset price and reflects the direction of the change (formal details are presented in section \ref{sec:metodology}). When constructed, negative realized semivariance ($RS^-$) and positive realized semivariance ($RS^+$) are able to measure volatility that accounts well for asymmetries in volatility evidenced on financial markets. Specifically, $RS^-$ has been shown to capture well the fact that future volatility depends more on past negative returns \citep{shephard2010measuring}.\footnote{In addition, negative jumps also contribute to future volatility more than their positive counterparts \citep{patton2011good}.}

\begin{figure}[h]
   \centering
   \includegraphics[width=5in]{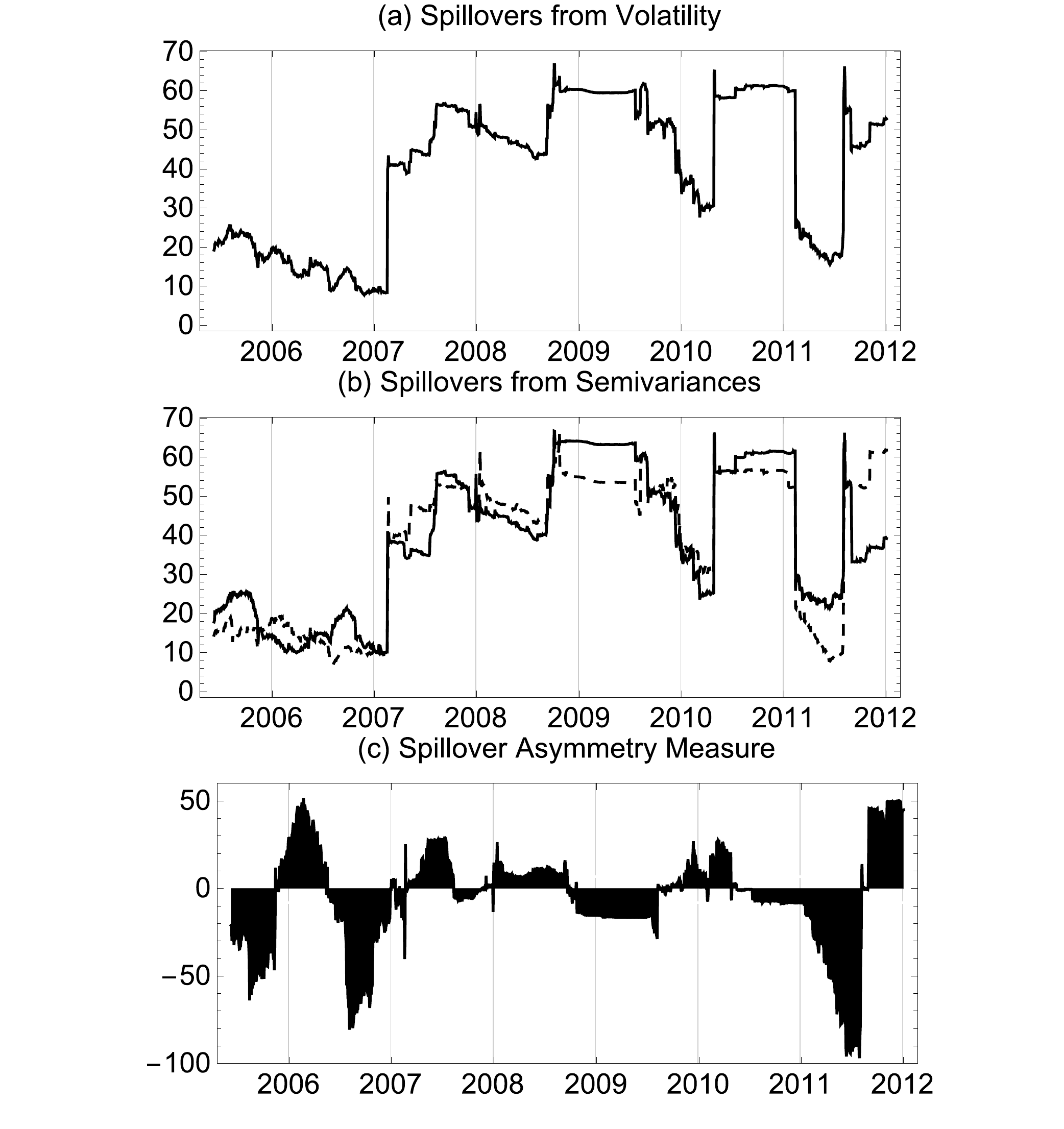} 
   \caption{(a) Spillovers from volatility (b) Spillovers from $RS^+$ (dashed), $RS^-$ (c) Spillover Asymmetry Measure ($\mathcal{SAM}$).} 
   \label{Fig:intro}
\end{figure}

We connect the two lines of research on volatility and its spillovers to assess our hypothesis that volatility spillovers exhibit asymmetries. Specifically, we extend the computation of the volatility spillover index due to \cite{diebold2012better} by allowing for negative and positive changes in returns to be considered separately via the realized semivariance due to \cite{shephard2010measuring}. As a result, by using negative realized semivariance ($RS^-$) and positive realized semivariance ($RS^+$) we are able to compute asymmetric volatility spillover indices robust to ordering in VAR. We illustrate our approach in Figure \ref{Fig:intro}. In panel (a), the total volatility spillovers based on the DY index are presented as they evolve over time. In panel (b), two separate lines represent volatility spillovers originating from bad and good volatility. Finally, in panel (c), the differences of both types of spillovers are pictured: spillovers due to bad volatility are captured in the negative domain and good volatility spillovers appear in the positive domain. The disparities are striking. What is more important is the fact that Figure \ref{Fig:intro} is not only an illustrative exercise but it shows the reality of volatility spillovers in the U.S. telecommunications services sector. Hence, disparities in volatility spillovers due to bad and good volatility do seriously matter as they produce markedly different effects over time. Therefore, we apply the method on 21 U.S. stocks with the highest market capitalization and show how the asymmetric volatility spillovers propagate across a wide range of different sectors.

The rest of the paper is organized as follows. In section \ref{sec:metodology} we formally introduce the concept of realized semivariance and the volatility spillover index and then describe the methodology for computing spillover indices on realized semivariance. In section \ref{sec:emp} we perform an empirical intra-market analysis. Section \ref{sec:conclus} briefly concludes.

\section{Measuring asymmetric volatility spillovers \label{sec:metodology}}

The main limitation of the current methodology for measuring spillovers from volatility is that it is used with measures of volatility based on daily (or weekly) high, low, opening, and closing prices. While the range-based volatility estimators are known to be good approximations of volatility, it is appealing to take advantage of the high-frequency data that are widely available nowadays and exploit the potential of recently developed realized volatility estimators. As we advocate in the introduction, it is interesting to use realized semivariance in the framework of spillover measures and obtain the spillovers due to volatility from both negative and positive returns.

In this section, we first introduce the two existing concepts, and then we describe a simple way how to combine them in order to capture asymmetric volatility spillovers using high-frequency measures.

\subsection{Realized variance and semivariance}

The first concept we introduce describes measures of volatility. Consider a continuous-time stochastic process for log-prices, $p_t$, evolving over a time horizon $[0\le t \le T]$, which consists of a continuous component and a pure jump component,
\begin{equation}
p_t=\int_0^t\mu_s ds + \int_0^t\sigma_s d W_s + J_t,
\end{equation}
where $\mu$ is a locally bounded predictable drift process and $\sigma$ is a strictly positive volatility process, and all is adapted to some common filtration $\mathcal{F}$. The quadratic variation of the log-prices $p_t$ is
\begin{equation}
[p_t,p_t] = \int_0^t\sigma_s^2 ds+\sum_{0<s\le t}(\Delta p_s)^2,
\end{equation}
where $\Delta p_s = p_s - p_{s-}$ are jumps, if present. A natural measure for quadratic variation has been formalized by \cite{andersen2001distribution} and \cite{barndorff2002econometric}, who propose estimating quadratic variation as the sum of squared returns and coined the name ``realized variance" ($RV$). Formally, let us suppose that the prices $p_0,\ldots,p_n$ are equally spaced on the interval $[0,t]$, then 
\begin{equation}
RV=\sum_{i=1}^n \left(p_i-p_{i-1}\right)^2
\end{equation}
converges in probability to $[p_t,p_t]$ with $n\rightarrow \infty$. More recently, \cite{shephard2010measuring} introduced estimators that capture the variation only due to negative or positive returns using an estimator of realized semivariance:
\begin{eqnarray}
RS^-&=& \sum_{i=1}^n r_i^2 I_{(r_i<0)} \\
RS^+&=& \sum_{i=1}^n r_i^2 I_{(r_i>0)}.
\end{eqnarray}
Realized semivariance provides a complete decomposition of the realized variance, as $RV=RS^- + RS^+$, and can serve as a measure of downside and upside risk. The decomposition holds exactly for any $n$. \cite{shephard2010measuring} show the limiting behavior of realized semivariance, which converges to $1/2\int_0^t\sigma_s^2 ds$ and the sum of the jumps due to negative and positive returns.

\subsection{Measuring volatility spillovers \label{sec:SI}}

The second concept we introduce describes how to measure volatility spillovers. \cite{diebold2009measuring} introduce a volatility spillover measure based on forecast error variance decompositions from vector auto regressions (VARs). Variance decompositions record how much of the $H$-step-ahead forecast error variance of some variable $i$ is due to innovations in another variable $j$, and hence provide a simple, intuitive way of measuring volatility spillovers. The methodology however has its limitations. First, it relies on the Cholesky-factor identification of VARs, and thus the resulting variance decompositions can be dependent on variable ordering. Second, a more crucial shortcoming of this methodology is that it allows measuring total spillovers only. Both limitations were successfully eliminated in a subsequent work. \cite{diebold2012better} use a generalized vector autoregressive framework in which forecast error variance decompositions are invariant to the variable ordering and that explicitly includes the possibility to measure directional volatility spillovers. 

Third, and most important to us, \cite{diebold2009measuring,diebold2012better} use the daily or weekly range-based volatility of \cite{garman1980estimation} to compute spillovers. Whereas range-based estimators provide an efficient way of estimating volatility, high-frequency data can further improve the understanding of the transmission mechanism. Due to the work of \cite{shephard2010measuring}, we can conveniently decompose daily volatility into negative (and positive) semivariance, providing a proxy for downside (and upside) risk. Replacing the total volatility, which enters the computation by the measures of downside (upside) risk, will allow us to measure the spillovers from the `bad' and `good' volatility, and test if they are transmitted in the same magnitude. Thus, we consider $\mathbf{RV_t} = (RV_{1t},\ldots,RV_{nt})'$ to measure total volatility spillovers, and $\mathbf{RS_t^{-}} = (RS^{-}_{1t},\ldots,RS^{-}_{nt})'$ and $\mathbf{RS_t^{+}} = (RS^{+}_{1t},\ldots,RS^{+}_{nt})'$ to measure volatility spillovers due to negative and positive returns, respectively.

To measure spillovers from the `bad' and `good' volatility, we use the \cite{diebold2012better} directional spillover measure, which follows directly from the variance decomposition associated with an $N$-variable vector autoregression fitted to volatility (in our case semivariances). To set the stage, consider an $N$-dimensional vector $\mathbf{RV_t} = (RV_{1t},\ldots,RV_{nt})'$ holding the realized variance of $N$ assets, which is modeled by a covariance stationary vector autoregression VAR($p$) as
\begin{equation}
\label{RV}
\mathbf{RV_t} = \sum_{i=1}^p \mathbf{\Phi}_i \mathbf{RV}_{t-i}+ \boldsymbol{\epsilon}_t,
\end{equation}
with $\boldsymbol{\epsilon}_t\sim N(0,\mathbf{\Sigma}_{\epsilon})$ being a vector of independently and identically distributed disturbances and $\mathbf{\Phi}_i$ for $i=1,\ldots,p$ coefficient matrices. Provided that the VAR process is invertible, it has the moving average representation 
\begin{equation}
\mathbf{RV}_t = \sum_{i=0}^{\infty}\mathbf{\Psi}_{i}\boldsymbol{\epsilon}_{t-i},
\end{equation}
where the $N\times N$ matrices holding coefficients $\mathbf{\Psi}_i$ can be obtained from the recursion $\mathbf{\Psi}_i = \sum_{j=1}^p\mathbf{\Phi}_j \mathbf{\Psi}_{i-j}$ with $\mathbf{\Psi}_0$ being the identity matrix; $\mathbf{\Psi}_0=\mathbf{I}_N$ and $\mathbf{\Psi}_i = 0$ for $i<0$. The moving average representation is key for understanding the dynamics of the system as it allows the computation of variance decompositions. These in turn allow the decomposition of the forecast error variances of each variable in the system into parts, which are attributable to various system shocks. \cite{diebold2012better} build the spillover index on the idea of assessing the fraction of the $H$-step-ahead error variance in forecasting the $i$th variable that is due to shocks to the $j$th variable for $j\ne i$, for each $i$. In order to obtain variance decompositions, which are invariant to variable ordering in the VAR system, \cite{diebold2012better} use the framework of the generalized VAR of \cite{koop1996impulse} and \cite{pesaran1998generalized}.\footnote{\cite{klossner2014exploring} developed a new algorithm for the fast calculation of the original DY index along with the computation of its minimum and maximum values.} The framework allows for correlated shocks but accounts for them by using the observed distribution of the errors, under a normality assumption. In this way, the shocks to each variable are not orthogonalized. Hence, the resulting sum of the contributions to the variance of the forecast error may not necessarily equal one.

\subsubsection{Total spillovers}
To define the total spillover index, \cite{diebold2012better} consider: (i) assets' own variance shares as the fractions of the $H$-step-ahead error variances in forecasting the $i$th variable that are due to assets' own shocks to $i$ for $i=1,\ldots,N$ and (ii) cross variance shares, or spillovers, as the fractions of the $H$-step-ahead error variances in forecasting the $i$th variable that are due to shocks to the $j$th variable, for $i,j=1,\ldots,N$, such that $i\ne j$. $H$-step-ahead generalized forecast error variance decomposition matrix $\Omega$ has following elements for $H=1,2,\ldots$
\begin{equation}
\omega_{ij}^H=\frac{\sigma_{jj}^{-1}\sum_{h=0}^{H-1}\left( \mathbf{e}'_i \mathbf{\Psi}_h \mathbf{\Sigma}_{\epsilon}\mathbf{e}_j \right)^2}{\sum_{h=0}^{H-1}\left( \mathbf{e}'_i \mathbf{\Psi}_h \mathbf{\Sigma}_{\epsilon}\mathbf{\Psi}'_h\mathbf{e}_i \right)},
\end{equation}
 where $\mathbf{\Sigma}_{\epsilon}$ is the variance matrix for the error vector, $\boldsymbol{\epsilon}_t$, $\sigma_{jj}$ is the standard deviation of the error term for the $j$th equation, $\mathbf{e}_i$ is the selection vector, with one as the $i$th element and zero otherwise, and $\mathbf{\Psi}_h$ are moving average coefficients from the forecast at time $t$. The sum of the elements in each row of the variance decomposition table is not equal to one, $\sum_{j=1}^N \omega_{ij}^H\ne1$, as the shocks are not necessarily orthogonal in this framework. Hence, we need to normalize each element by the row sum as:
 \begin{equation}
 \widetilde{\omega}_{ij}^H = \frac{\omega_{ij}^H}{\sum_{j=1}^N \omega_{ij}^H}.
 \end{equation} 
Using the contributions from the variance decomposition, \cite{diebold2012better} then define the total spillover index, which measures the contribution of spillovers from volatility shocks across variables in the system to the total forecast error variance as
\begin{equation}
\label{stot}
\mathcal{S}^H=100\times \frac{1}{N} \sum_{\substack{i,j=1\\ i\ne j}}^N\widetilde{\omega}_{ij}^H.
\end{equation}
Note that by construction, $\sum_{j=1}^N \widetilde{\omega}_{ij}^H=1$ and $\sum_{i,j=1}^N \widetilde{\omega}_{ij}^H=N$, thus the contributions of spillovers from volatility shocks are normalized by the total forecast error variance.

\subsubsection{Directional spillovers}

The total spillover index as defined by equation (\ref{stot}) helps us understand how much of the shocks to volatility spill over across the studied assets. However, the main advantage of the generalized VAR framework is its ability to identify directional spillovers using the normalized elements of the generalized variance decomposition matrix. Directional spillovers allow us to further uncover the transmission mechanism, as we can decompose the total spillovers to those coming from, or to, a particular asset in the system.

\cite{diebold2012better} propose to measure the directional spillovers received by asset $i$ from all other assets $j$ as:
\begin{equation}
\mathcal{S}_{i\leftarrow\bullet}^H=100\times \frac{1}{N} \sum_{\substack{j=1\\ i\ne j}}^N\widetilde{\omega}_{ij}^H.
\end{equation}
In a similar fashion, the directional spillovers transmitted by asset $i$ to all other assets $j$ can be measured as:
\begin{equation}
\mathcal{S}_{i\rightarrow\bullet }^H=100\times \frac{1}{N} \sum_{\substack{j=1\\ i\ne j}}^N\widetilde{\omega}_{ji}^H.
\end{equation}

\subsubsection{Net spillovers and net pairwise spillovers}
Directional spillovers can be also used to obtain the net volatility spillover from asset $i$ to all other assets $j$. The directional spillover is then defined as the simple difference between gross volatility shocks transmitted to and received from all other assets:
\begin{equation}
\mathcal{S}^H_i=\mathcal{S}_{i\rightarrow\bullet }^H-\mathcal{S}_{i\leftarrow\bullet}^H.
\end{equation}
The net volatility spillover tells us how much each asset contributes to the volatility in other assets in net terms. 

Finally, the pairwise volatility spillover between asset $i$ and $j$ can be simply defined as the difference between the gross shocks transmitted from asset $i$ to asset $j$ and those transmitted from $j$ to $i$:
\begin{equation}
\mathcal{S}^H_{ij}=100\times \frac{1}{N} \left(\widetilde{\omega}_{ji}^H-\widetilde{\omega}_{ij}^H \right).
\end{equation}

\subsection{Measuring asymmetric spillovers \label{sec:MAS}}

We now describe how to capture and measure asymmetric volatility spillovers. Specifically, we are able to account for spillovers from volatility due to negative returns ($\mathcal{S}^-$) and positive returns ($\mathcal{S}^+$), as well as directional spillovers from volatility due to negative returns ($\mathcal{S}_{i\leftarrow\bullet}^-$, $\mathcal{S}_{i\rightarrow\bullet}^-$), and positive returns ($\mathcal{S}_{i\leftarrow\bullet}^+$, $\mathcal{S}_{i\rightarrow\bullet}^+$). Based on the previous exposition, to isolate asymmetric volatility spillovers we need to replace the vector of volatilities $\mathbf{RV_t} = (RV_{1t},\ldots,RV_{nt})'$ defined in equation (\ref{RV}) with the vector of negative semivariances $\mathbf{RS_t^{-}} = (RS^{-}_{1t},\ldots,RS^{-}_{nt})'$ or the vector of positive semivariances $\mathbf{RS_t^{+}} = (RS^{+}_{1t},\ldots,RS^{+}_{nt})'$. Note that in the above definitions we dropped the $H$ index to ease the notational burden, but it remains a valid parameter for the estimation of spillover indices.

Spillovers due to negative and positive returns now enable analyzing the effect of bad and good volatility. A comparison of the spillover values opens the following possibilities. If the contributions of $RS^-$ and $RS^+$ are equal, the spillovers are symmetric. On the other hand, the differences in the realized semivariances result in asymmetric spillovers. These properties enable us to test the following hypotheses.
$$\begin{array}{ccccccc}
H_0: && \mathcal{S}^- = \mathcal{S}^+ & \text{against} & H_A: && \mathcal{S}^- \ne \mathcal{S}^+\\
H_0: &&\mathcal{S}_{i\leftarrow\bullet}^- = \mathcal{S}_{i\leftarrow\bullet}^+ & \text{against} & H_A: && \mathcal{S}_{i\leftarrow\bullet}^- \ne \mathcal{S}_{i\leftarrow\bullet}^+\\
H_0: && \mathcal{S}_{i\rightarrow\bullet}^- = \mathcal{S}_{i\rightarrow\bullet}^+& \text{against} & H_A: &&  \mathcal{S}_{i\rightarrow\bullet}^- \ne \mathcal{S}_{i\rightarrow\bullet}^+
\end{array}$$
Rejecting a null hypothesis means that bad and good volatility does matter for spillover transmission in terms of magnitude as well as direction. Moreover, we assume that the values of the volatility spillover indices differ over time. To capture the time-varying nature, we compute the indices using a moving window.

\subsubsection{Spillover Asymmetry Measure}
\label{samsec}
In order to better quantify the extent of volatility spillovers we introduce a spillover asymmetry measure ($\mathcal{SAM}$) that is formally defined as
\begin{equation}
\label{sam}
\mathcal{SAM} = 100\times \frac{\mathcal{S}^+-\mathcal{S}^-}{1/2 \left(\mathcal{S}^++\mathcal{S}^-\right)},
\end{equation}
where $\mathcal{S}^-$ and $\mathcal{S}^+$ are volatility spillover indices due to negative and positive semivariances, $RS^-$ and $RS^+$, respectively, with an $H$-step-ahead forecast at time $t$. $\mathcal{SAM}$ defines and illustrates the extent of asymmetry in spillovers due to $RS^-$ and $RS^+$. When $\mathcal{SAM}$ takes the value of zero, spillovers coming from $RS^-$ and $RS^+$ are equal. When $\mathcal{SAM}$ is positive, spillovers coming from $RS^+$ are larger than those from $RS^-$ and the opposite is true when $\mathcal{SAM}$ is negative.

\subsubsection{Directional Spillover Asymmetry Measure}
While the spillover asymmetry measure ($\mathcal{SAM}$) defined by equation (\ref{sam}) measures to what extent the spillovers from volatility are asymmetric, we can decompose this measure and study the source of asymmetry among the studied assets. 
We define the asymmetry measure for directional spillovers received by asset $i$ from all other assets $j$ as
\begin{equation}
\mathcal{SAM}_{i\leftarrow\bullet} = 100\times \frac{\mathcal{S}_{i\leftarrow\bullet}^+-\mathcal{S}_{i\leftarrow\bullet}^-}{1/2 \left(\mathcal{S}_{i\leftarrow\bullet}^++\mathcal{S}_{i\leftarrow\bullet}^-\right)}.
\end{equation}
In a similar fashion, we can measure the degree of asymmetry in directional spillovers transmitted by asset $i$ to all other assets $j$:
\begin{equation}
\mathcal{SAM}_{i\rightarrow\bullet} = 100\times \frac{\mathcal{S}_{i\rightarrow\bullet}^+-\mathcal{S}_{i\rightarrow\bullet}^-}{1/2 \left(\mathcal{S}_{i\rightarrow\bullet}^++\mathcal{S}_{i\rightarrow\bullet}^-\right)}.
\end{equation}
$\mathcal{SAM}_{i\leftarrow\bullet}$ and $\mathcal{SAM}_{i\rightarrow\bullet}$ allow us to identify the extent to which volatility from (or to) the $i$th asset spills over to (or from) other assets symmetrically. For example, if `bad' volatility from one asset in the system transmits to volatilities of other assets more than the `good' volatility, $\mathcal{SAM}_{i\rightarrow\bullet}$ will be different from zero, and we expect it to be negative. This information would stay hidden in the original \cite{diebold2012better} framework.

\subsection{Bootstrapping $\mathcal{SAM}$}
The definitions of $\mathcal{SAM}$, $\mathcal{SAM}_{i\leftarrow\bullet}$ and $\mathcal{SAM}_{i\rightarrow\bullet}$ help us to test our earlier hypotheses about the symmetry of spillovers. When using these spillover asymmetry measures, the former three hypotheses are re-defined in the following way:
$$\begin{array}{ccccccc}
H_0: && \mathcal{SAM}=0 & \text{against} & H_A: && \mathcal{SAM} \ne 0\\
H_0: &&\mathcal{SAM}_{i\leftarrow\bullet} = 0 & \text{against}& H_A: && \mathcal{SAM}_{i\leftarrow\bullet} \ne 0\\
H_0: && \mathcal{SAM}_{i\rightarrow\bullet} = 0 & \text{against}& H_A: &&  \mathcal{SAM}_{i\rightarrow\bullet} \ne 0.
\end{array}$$
In case we reject one of the null hypotheses against the alternative, we are also able to distinguish to what extent the `bad' and `good' volatility spills over. This is done based on the sign of $\mathcal{SAM}$, $\mathcal{SAM}_{i\leftarrow\bullet}$, and $\mathcal{SAM}_{i\rightarrow\bullet}$. If there is a specific version of $\mathcal{SAM}<0$, the `bad' volatility spills more than the `good' one, and vice versa for $\mathcal{SAM}>0$.

In order to test the above hypotheses about the symmetry of volatility spillovers, we opt for bootstrapping the measures. It is important to confirm that the empirical results in further sections are not possible due to the estimation error from generalized VAR, or the discretization error from realized semivariances. Especially the latter could be high due to a small number of observations during the day in the real data used for the computation of realized semivariances. Hence we simulate the stochastic volatility processes, which will be connected via simple correlation, but will contain no asymmetry nor other dependencies. In the bootstrapping procedure, we will set the parameters as close as possible to the real world data used later in the study. Hence we can use the simulations to test the small sample performance of the proposed asymmetric spillover indices. 

Using the setup of \cite{BN2011}, we simulate a bivariate factor stochastic volatility model for $X_{i,t}$, $i=\{1,2\}$ and $t\in[0,1]$ as:
\begin{eqnarray}
\label{MCmodel2}
\nonumber dX_{i,t} &= &\mu_{i}dt+\gamma_i \sigma_{i,t} dB_{i,t}+\sqrt{1-\gamma_i^2} \sigma_{i,t} dW_{t}+c_{i,t} dN_{i,t}\\
\nonumber d\sigma_{i,t} &=&\exp(\beta_0+\beta_1 v_{i,t}) \\
dv_{i,t}&=&\alpha v_{i,t} dt+dB_{i,t},
\end{eqnarray}
where the elements of $B_{i,t}$ are independent standard Brownian motions and are also independent of $W_t$, and $c_{i,t} dN_{i,t}$ are independent compound Poisson processes with random jump sizes distributed as $N\sim(0,\sigma_{1})$. We simulate the processes using the Euler discretization scheme at a time interval of $\delta=1s$, each with $6.5\times60\times60$ steps ($n=23,400$), corresponding to a 6.5-hour trading day. Following \cite{BN2011}, the parameters are set to $(\mu_1,\mu_2,\beta_0,\beta_1,\alpha,\gamma_1,\gamma_2)=(0,0,-5/16,1/8,-1/40,-0.3,-0.3)$. Each day is restarted with an initial value of $v_{i,t}$ drawn from a normal distribution $N(0,(-2\alpha)^{-1})$. On each simulated path, we estimate RV, $RS^+$, and $RS^-$ using 5-minute returns so we uncover a possible error due to small samples in the empirical data. Then, the  $\mathcal{SAM}$, $\mathcal{SAM}_{i\leftarrow\bullet}$, and $\mathcal{SAM}_{i\rightarrow\bullet}$ are estimated. 

The true spot correlation between $X_{1,t}$ and $X_{2,t}$ is set to $\sqrt{(1-\gamma_1^2)(1-\gamma_2^2)}$, which is equal to 0.91; hence, processes are strongly correlated and reflect the high correlation found in the empirical variances. We set the sampling of realized measures to 5 minutes, being close to the average of 36 observations, to compute realized semivariance per day. Using 10,000 simulated processes, we obtain the average value of SAM together with the 95\% confidence intervals computed as quantiles from simulated data: $(-6.6728, -0.0342, 6.7650)$. Hence the expected value of the $\mathcal{SAM}$ measure is not statistically different from zero, and we can use the simulated confidence intervals to test our hypotheses on real world data.

\section{Asymmetric connectedness of U.S. stocks \label{sec:emp}}

Using the proposed methodology, we study the connectedness of the 21 most liquid U.S. stocks from the seven main market sectors defined in accordance with the Global Industry Classification Standard (GICS).\footnote{The Global Industry Classification Standard is an industry taxonomy developed by MSCI and Standard \& Poor's for use by the global financial community.} The three stocks with the highest market capitalization in a sector define the analyzed sector.\footnote{Those three stocks account for approximately half of the total capitalization of the sector. Note that this is an approximate extent that varies over time.} All sectors and stocks are listed in Table \ref{tab:sectors}. We cover the period from August 2004 to December 2011. The period under study is very informative in terms of market development, sentiment, and expectations since we cover the 2007--2008 financial crisis plus three years before and after the crisis. The data were obtained from the Price-Data.com.\footnote{\url{http://www.price-data.com/}}

For the computation of realized measures, we restrict the analysis to 5-minute returns during the 9:30 a.m. to 4:00 p.m. business hours of the New York Stock Exchange (NYSE). The data are time-synchronized by the same time-stamps, eliminating transactions executed on Saturdays and Sundays, U.S. federal holidays, December 24 to 26, and December 31 to January 2, because of the low activity on these days, which could lead to estimation bias. Hence, we work with data from 1835 trading days.  
\begin{table}
\begin{tabular}{lcp{10cm}}
\toprule
Sector&& Stocks \\
\midrule
Financials && Bank of America Corporation (BAC), Citigroup (C), Wells Fargo \& Company (WFC) \\
Information Technology && Apple (AAPL), Intel Corporation (INTC), Microsoft Corporation (MSFT)\\
Energy && Chevron Corporation (CVX), Schlumberger Limited (SLB), Exxon Mobil Corporation (XOM)\\
Consumer Discretionary && Amazon.com (AMZN), Walt Disney Company (DIS), McDonaldÕs Corp. (MCD)\\
Consumer Staples && Coca-Cola Company (KO), Procter \& Gamble Co. (PG), Wal-Mart Stores (WMT)\\
Telecommunication Services && Comcast Corporation (CMCSA), AT\&T (T), Verizon Communications (VZ)\\
Health Care && Johnson \& Johnson (JNJ), Merck \& Co. (MRK), Pfizer (PFE) \\
\bottomrule
\end{tabular}
\caption{List of stocks used in sectors.}
\label{tab:sectors}
\end{table} 
Using a large number of stocks from the liquid U.S. market, we perform an intra-market analysis on a homogeneous class of assets, as suggested in \cite{diebold2009measuring}, and differ from much of the empirical research that concentrates on spillovers among international markets. 

\subsection{Intra-market connectedness}

Before studying asymmetries, we first inspect the overall extent of U.S. intra-market connectedness. The recent 2008 crisis influenced all sectors and changed the way market participants as well as ordinary citizens perceive risk \citep{burns2012}. Hence, it may be logical to expect an increased post-crisis connectedness of stocks due to more homogenous beliefs of market participants, who expect higher levels of risk. Volatility, as a measure of risk in markets, can help us to uncover to what extent this sharp change in expectations influences also the connectedness of stock markets. 

\begin{figure}[h]
   \centering
   \includegraphics[width=4in]{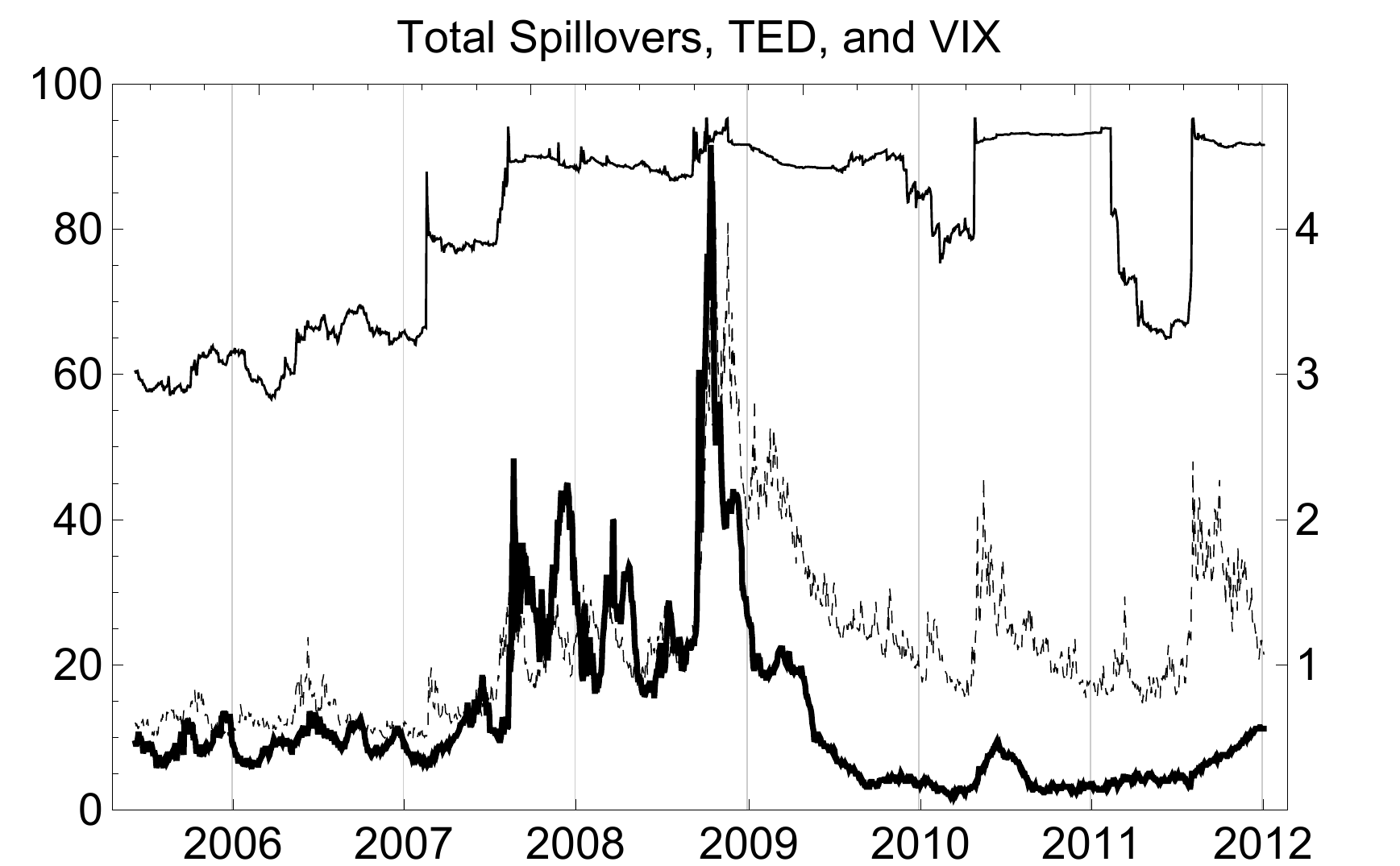}    
   \caption{Spillovers from volatility in contrast to TED spread (bold) and VIX index (dashed).}
   \label{Fig:tedvix}
\end{figure}

The volatility spillover index computed for the portfolio of 21 stocks under research is depicted in Figure \ref{Fig:tedvix}. In order to capture the dynamics of spillovers, we use a rolling window of length 200.\footnote{Any point of the volatility spillover index at time $t$ is computed on the interval $(t-199,t)$.} Since we hypothesize that the level of spillovers is connected to the level of market risk, we contrast the spillovers to two proxies of risk in Figure \ref{Fig:tedvix}. First, we use the VIX index,\footnote{The Chicago Board Options Exchange Market Volatility Index represents a measure of the market's expectation of stock market volatility over the next 30 day period.} well known among investors as a ``fear index": an increasing VIX index means that market participants expect an increase in market volatility. Second, we use the TED spread\footnote{The TED spread is calculated as the difference between the three-month LIBOR and the three-month U.S. T-bill interest rate. TED is an acronym formed from T-Bill and ED, the ticker symbol for the Eurodollar futures contract.} as a general indicator of perceived credit risk in the economy. An increase (decrease) in the TED spread is a sign that lenders believe the risk of default on interbank loans is increasing (decreasing).

The level of volatility spillovers for our aggregate portfolio is relatively low at the beginning of the studied period, whereas the spillovers increase substantially during the active phase of the crisis. The first substantial increase in stock market connectedness is detected on August 2007, when BNP Paribas terminated withdrawals from three hedge funds due to ``a complete evaporation of liquidity in certain market segments of the U.S. securitisation market" (\citeauthor{HowardGreen2010}, 2010; p.1). The connectedness of U.S. stocks reached a maximum around 90\% after the crisis's peak in September and October 2008.

To sum up, the overall intra-market connectedness of the U.S. stocks increased substantially with the increasing uncertainty of stock market participants, highlighted by the level shift in the ex-ante volatility from VIX index as well as the TED spread.

\subsubsection{Connectedness at the disaggregate sectoral level}

\begin{figure}[h]
   \centering
   \includegraphics[width=2.3in]{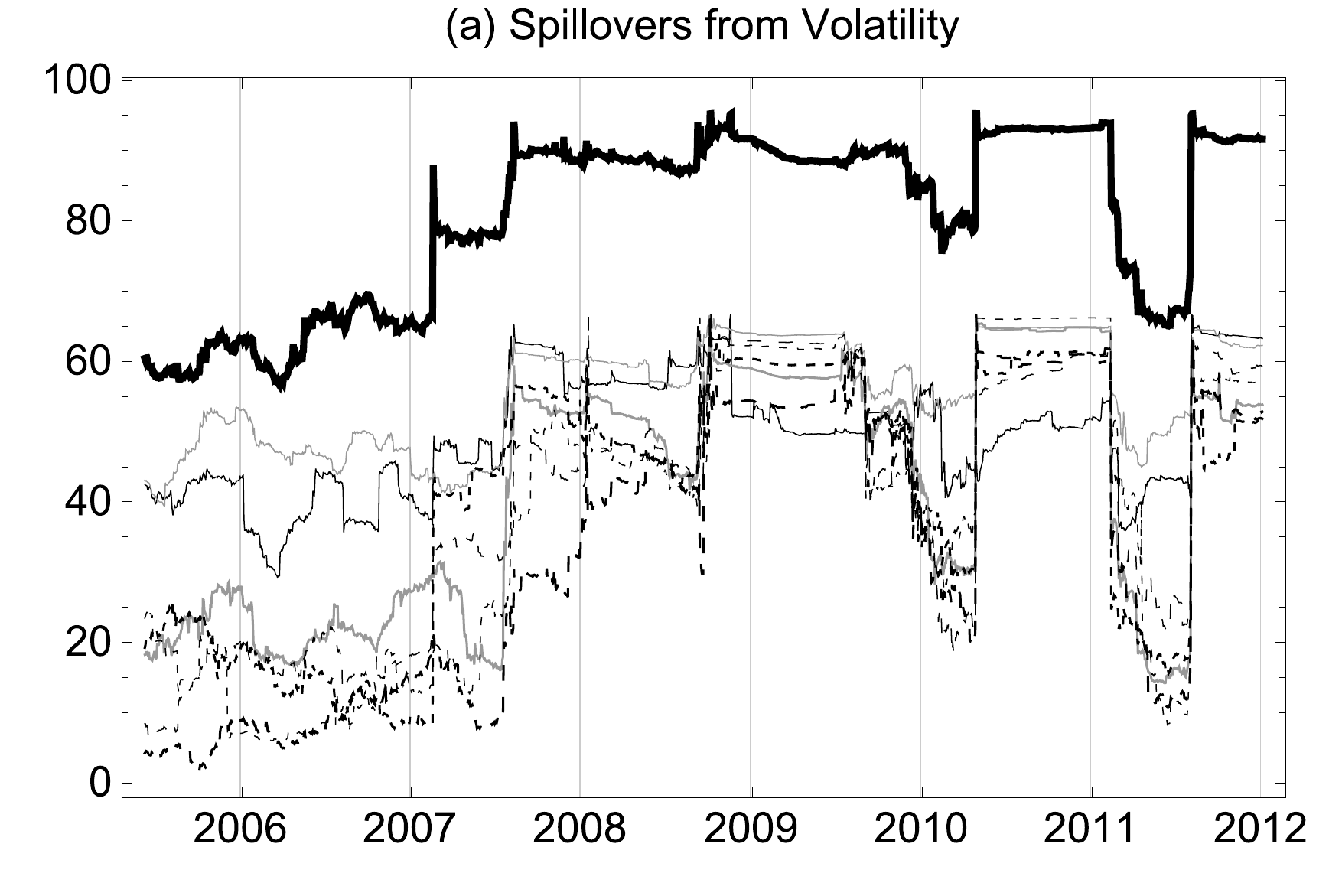} 
      \includegraphics[width=3.45in]{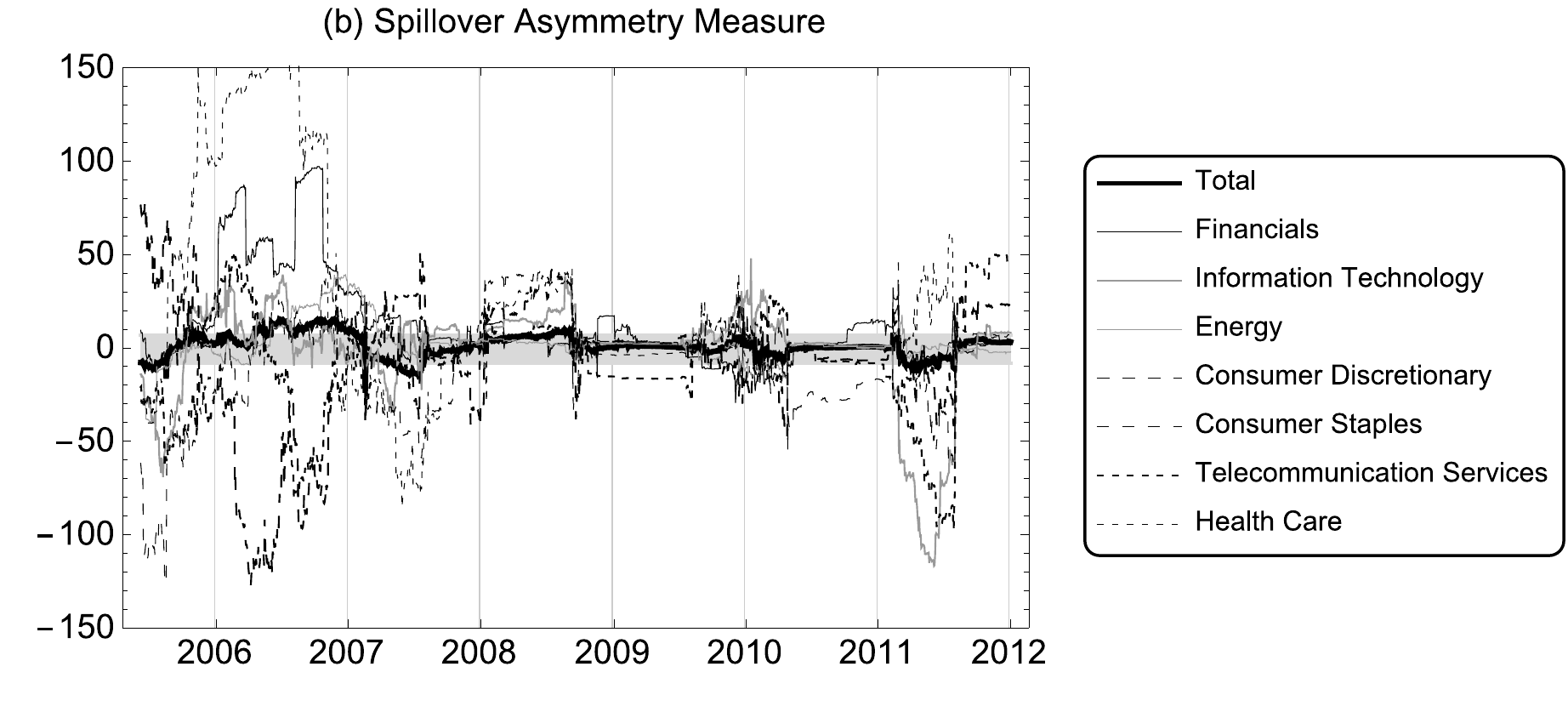} 
   \caption{(a) Spillovers from volatility (b) Spillover Asymmetry Measure ($\mathcal{SAM}$)} comparison for all sectors as well as total.
   \label{Fig:comparison}
\end{figure}

In order to better understand how volatility spills over among stocks within a portfolio, we compare the volatility spillovers for separate sectors. Figure \ref{Fig:comparison} shows the dynamics of the sectoral volatility spillovers. While the impact of increased uncertainty marked by the Lehman Brothers collapse in September 2008 is visible for all sectors in the substantial increase of the market connectedness, another observation emerges as well. 

The connectedness of stocks in the financial and energy sectors is two times as large during the 2005--2007 period when compared to the rest of sectors. It is useful to turn our attention to the TED spread, which serves as a reliable indicator of uncertainty. A rising TED spread indicates that liquidity is being withdrawn from the market and the withdrawal can be perceived as an increase in uncertainty. The first significant increase of the TED spread occurs in August 2007 (from 50 to 200 bps) while the second increase occurs in September 2008 when the TED spread jumped to values over 450 bps. These substantial increases are strongly correlated with an increase in total volatility spillovers for the whole portfolio as well as for each sector. Hence, we can draw a conclusion that volatility spillovers are significantly higher in periods when the market is under pressure and experiences homogenous expectations about rising risk. During August 2007 and September 2008, sectoral spillovers of the two groups synchronize and point towards more homogeneous expectations of market participants about risk level. 

During the second half of 2008 and early 2009, we see a significant decrease of the TED spread, but volatility spillovers remain at the same high level. It indicates that market uncertainty, translated to stock connectedness, has not decreased. After 2008 we notice two synchronized declines in sectoral volatility spillovers. These declines always end with a sharp increase of the VIX index. After 2008 the high spillovers are reflected in high levels of the VIX but the levels of the TED decline considerably. The reason for the difference between the TED and spillovers stems from the dynamics of the two components that form the TED spread. The three-month T-bill interest rate fell considerably during 2008 and since 2009 it remained close to zero. The three-month LIBOR rate followed a similar path and during 2010--2011 hovered in the vicinity of 30 bps. Hence, the room for the spread calculated from the two components remained very thin and the low level shown earlier in Figure \ref{Fig:tedvix} is therefore in sharp contrast to high spillovers.

What is the economic rationale behind the observed evidence? Volatility is generally viewed as a risk measure: low (high) volatility implies low (high) risk. Hence, volatility spillovers can be perceived as risk spillovers. Volatility is also related to the rate of information flow in a market \citep{ross1989information} and information flow is key for risk management in general. Hence, understanding how volatility evolves and is transmitted (spills over) is important to monitor the current state of the market and to provide indications of future risk development, risk assessment, and related decisions. 

A preliminary analysis of intra-market connectedness shows that sectoral volatility spillovers clearly exhibit a structural change induced by the crisis. Furthermore, the portfolio disaggregation reveals that values of volatility spillovers significantly differ across sectors. This dynamics can be attributed to the fact that expectations of increased risk were transferred to the higher connectedness among U.S. stocks. As indicated earlier, we hypothesize that it matters whether the spillovers are due to bad or good volatility. We analyze this issue in the next section.

\subsection{How does bad and good volatility spill over U.S. sectors?}

The previous analysis was concerned with the overall risk level and its transmission among stocks. While the overall level of U.S.-stock connectedness is high, we may believe that risk due to negative and positive news is perceived differently by market participants and spillovers would exhibit asymmetries. Then, it is interesting to see whether the different perception of risk is also transmitted across stocks. The decomposition of volatility into bad and good volatility based on negative and positive returns can be perceived as a degree of downside and upside risk \citep{fenou2013}. Bad volatility may result from a single negative news of high importance, increased political risk, worsening of the economic conditions, etc.; a common feature of bad volatility shocks is that they are prone to dampen economic growth or decrease shareholdersÕ value. Good volatility may also materialize for various reasons: due to positive macroeconomic, sectoral, or firm-specific announcements; legal or tax measures improving business conditions; the end of a recession; etc. More details on bad and good volatility are provided, for example, in \cite{bartram2012us,fenou2013} or \cite{Segal2014}.

\subsubsection{Asymmetric connectedness and aggregation}
We study volatility transmission asymmetry with the help of our spillover asymmetry measure ($\mathcal{SAM}$), defined in section \ref{samsec}. To be able to test the null hypothesis of symmetrical connectedness, we use bootstrap confidence intervals constructed as described in section 2.4. The $\mathcal{SAM}$ for the whole portfolio of 21 stocks reveals a rather surprising outcome. There is almost no difference in volatility spillovers due to bad or good volatility (see Figure \ref{Fig:comparison}b), as we cannot reject the null hypothesis that $\mathcal{SAM}=0$. This result indicates that the portfolio is well balanced in the sense that effects of good and bad volatility are more or less equal. It also indicates that asymmetry is not present at the aggregate portfolio level, which contradicts findings from the previous literature. 

\cite{bekaert2000asymmetric} document asymmetry in aggregate market returns. The authors argue that part of the reason for the asymmetry in aggregate market returns, as well as an explanation for the aggregate returns being more asymmetric than firm level returns, is that firm returns are more correlated in downside markets. Hence, when the market declines, the aggregate stock market volatility rises as covariance rises. In addition, \cite{avramov2006impact} argue that a positive return is followed by selling activity that is dominated by informed traders who tend to reduce volatility. On the contrary, a negative return is followed by selling activity that is dominated by uninformed traders who tend to increase volatility. Based on this literature, we may speculate that an increase in negative volatility due to uninformed investors in one stock may lead to an increase in volatility in other stocks as well.

In contrast to the documented asymmetry at the disaggregated level, the spillovers in a larger portfolio of stocks are symmetric. The main reason for this may come from firm-level or sector-level heterogeneity, as opposed to symmetry at the aggregate level. Therefore, we turn our attention to study asymmetries on the sectoral level by analyzing the transmission mechanism of bad and good volatility in portfolios containing stocks from seven specific sectors.

\subsubsection{Asymmetric connectedness of stocks at the sectoral level}

We employ $\mathcal{SAM}$ to assess spillovers within individual sectors. The disaggregation of sectors shows a striking result. Figure \ref{Fig:comparison}b reveals that stocks at sectoral levels exhibit substantial asymmetries in volatility spillovers and the asymmetries differ across sectors.

How should we interpret the empirically found asymmetries? Recall that the $\mathcal{SAM}$ in panel (b) is constructed as the moving window of the differences between bad and good volatility spillovers. The observed clusters of these differences actually show that asymmetries in spillovers have a tendency to propagate in a specific domain for some period of time. Hence, the dynamics of $\mathcal{SAM}$ shows whether positive or negative spillovers dominate and can serve as an approximation of positive or negative market expectations about present and near-future market movements. In fact, $\mathcal{SAM}$ measures how the (majority of agents on) markets are sensitive to bad or good news, and how the news - which prompts bad and good volatility - transmits within sectoral portfolios of stocks. Therefore, $\mathcal{SAM}$ can also be a good indicator of whether the market is in an optimistic or pessimistic mood and what are the expectations.

As to the economic interpretation, we document the sector-level heterogeneity in the transmission mechanism. There are periods when informed traders dominating the market tend to reduce volatility, and this good volatility is then transmitted at a much higher magnitude to other stocks within the sector than bad volatility ($\mathcal{SAM}>0$). On the other hand, there are also periods when uninformed traders dominate the market and bad volatility tends to transmit at a much larger magnitude ($\mathcal{SAM}<0$). 

\begin{figure}[h]
   \centering
   \includegraphics[width=7in]{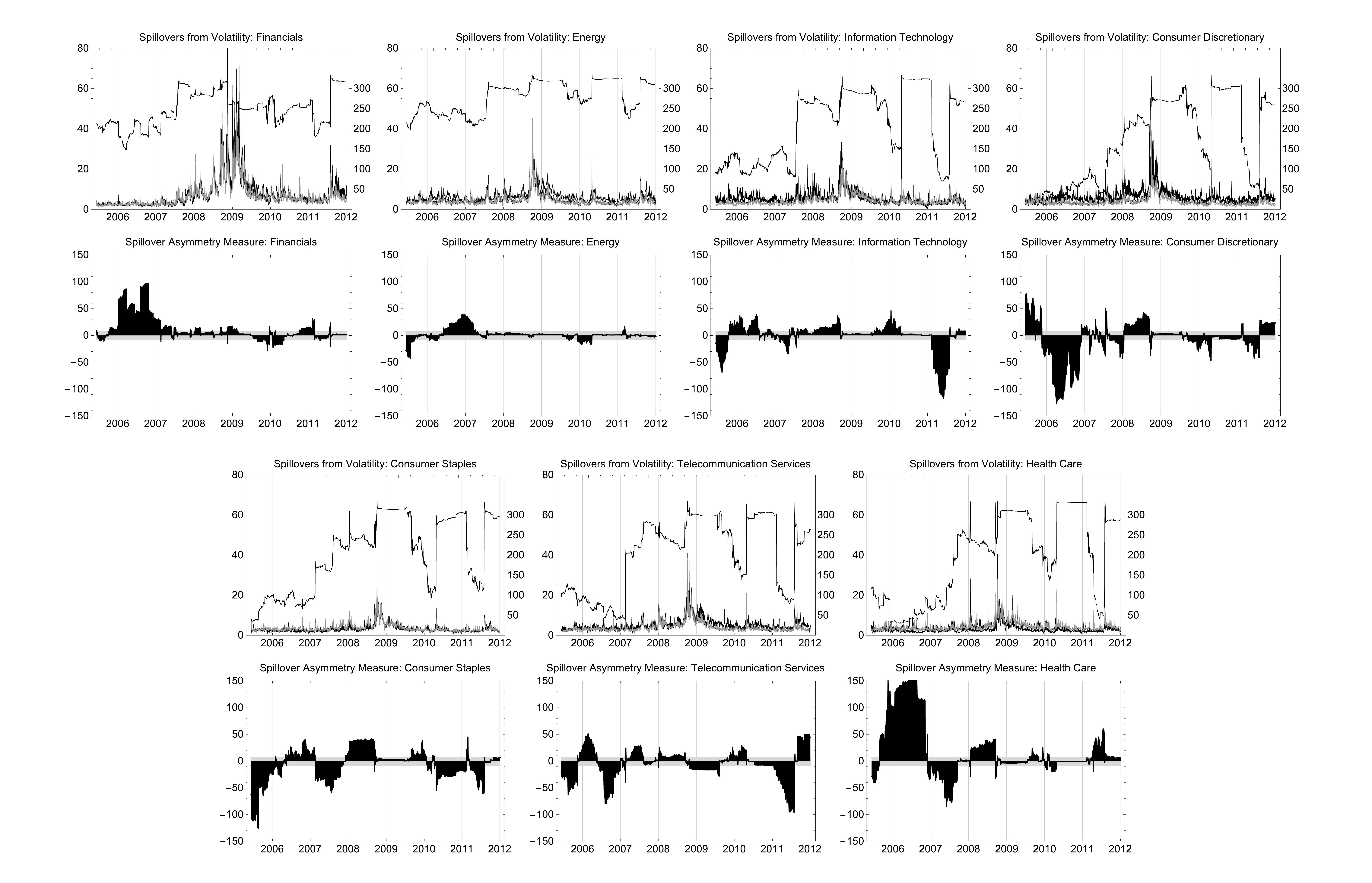} 
   \caption{Realized volatility, spillover index (upper panels), and Spillover Asymmetry Measure, $\mathcal{SAM}$ (lower panels) for all sectors.}    
   \label{Fig:SAMfin}
\end{figure}

In order to closely inspect all the studied sectors, and to present the extent of asymmetries together with total spillovers and the volatility of individual stocks, we present Figure \ref{Fig:SAMfin}, where each sector is represented by two panels. The upper panel depicts the total volatility spillover and realized volatility of each stock in the sector; the lower panel displays $\mathcal{SAM}$. In order to provide an easier comparison between sectors, we use an equal scale on each vertical axis. Indeed, the volatility of individual stocks plays an important role in the spillover analysis. In almost all sectors, the increase of volatility spillovers corresponds to sudden increases of volatility levels of individual stocks. This feature can be easily found in the upper panels for specific sectors (Figure \ref{Fig:SAMfin}). Recall that the relation between volatility and the spillover index is qualitatively similar for the aggregated total spillover index and the VIX index (see Figure \ref{Fig:tedvix}). Based on the evidence presented in Figure \ref{Fig:SAMfin}, we discuss the results for individual sectors below.

The financial sector exhibits the second-highest extent of total spillovers prior to the financial crisis as well as high magnitudes thereafter. $\mathcal{SAM}$ records a substantial contribution of spillovers due to positive returns during the pre-crisis period with the peak coinciding with the sub-prime crisis in late 2006. The large proportion of positive asymmetry before the 2007--2008 financial crisis is the second largest among the sectors under research (the health care sector records even larger positive asymmetries). The evidence is in line with the findings of \cite{bartram2012us}, who show that higher volatility in U.S. firms is due to good volatility resulting from conditions conducive to growth; prior to the financial crisis the financial sector was thriving on foreign money inflows, low interest rates, and real-estate overpricing \citep{shiller2012subprime}. A sharp decline in asymmetric spillovers follows after the crisis erupted. Since 2009 the pattern of alternating positive and negative asymmetries prevails. However, when analyzing Figure \ref{Fig:SAMfin} at the full length of the time axis, an optimistic mood prevails on the market.

When compared to other sectors, the energy sector is exceedingly symmetric in terms of good and bad volatility spillovers, despite the fact that this sector records the highest extent of total spillovers. The defensive nature of the sector paired with the relative price stability of the stocks should be a reason behind the low asymmetry in spillovers. Still, stock prices in the energy sector reflect the sharp increase in the price of oil in the last half of 2007 and first half of 2008 that could well be fueled by increased speculation and heightened activity in commodity markets. Positive values of $\mathcal{SAM}$ visible in Figure \ref{Fig:SAMfin} during exactly the same period are typical examples of spillovers due to positive returns. A much smaller spike in positive $\mathcal{SAM}$ can be spotted in early 2011 when political turmoil in Egypt, Libya, Yemen, and Bahrain drove oil prices to \$95 per barrel in late February 2011 and even to \$103 per barrel by February 24, 2011 when oil production was curtailed by the political upheaval in Libya. Hence, the asymmetries in spillovers do occur in the energy sector but with the exception of the pre-crisis period they are quite small. 

The information technology sector exhibits strong negative as well as positive asymmetries before and during the crisis. The post-crisis decline in asymmetries is replaced by extremely negative asymmetric spillovers during the first half of 2011. However, with the exception of the two large negative asymmetry clusters in 2005 and 2011, the spillovers in the technology sector can be characterized as driven primarily by good volatility.

Consumer discretionary is a sector where the distinction between bad and good volatility spillovers becomes critical. Substantial but declining positive asymmetries in spillovers in the beginning of the period under research are converted into massive negative spillovers during 2006, just before the financial crisis began. The crisis period itself is characterized by an alternating pattern of positive and negative spillovers but the effect of bad volatility is clearly visible and a pessimistic mood dominates most of the analyzed period.

The defensive nature of the consumer staples sector is in contrast to the propagation of asymmetries, which are mainly due to negative spillovers. Asymmetries in spillovers in the whole sector are disproportionally larger than positive ones during both pre-crisis and post-crisis periods. The clusters of negative spillovers document the pessimistic mood in the sector that under normal circumstances should provide balance to a portfolio and a low-risk profile with respect to a falling market. Despite the abrupt changes in prices indicated above, the relative stability of prices is confronted with numerous negative risk spillovers that do not resonate well with the expectations of future stability in this particular sector.

In terms of asymmetries, the telecommunication services sector offers a quite regular oscillatory pattern of asymmetric spillovers and the clusters do not exhibit any clear domination due to bad or good volatility. Finally, health care exhibits the largest extent of positive spillovers among all sectors. In the pre-crisis period, the difference in spillover asymmetry measured by $\mathcal{SAM}$ reaches over 150\%. The pattern does not relate solely to the pre-crisis period as spillovers due to positive volatility dominate in general. 

To conclude, we confirm a clear asymmetric connectedness of markets at the disaggregate sectoral level, which is in contrast to the symmetric volatility transmission mechanism at the aggregate level. The result can be attributed to large sector-level heterogeneity.

\subsection{Contribution of individual stocks to asymmetry}

\begin{figure}[h]
   \centering
   \includegraphics[width=5in]{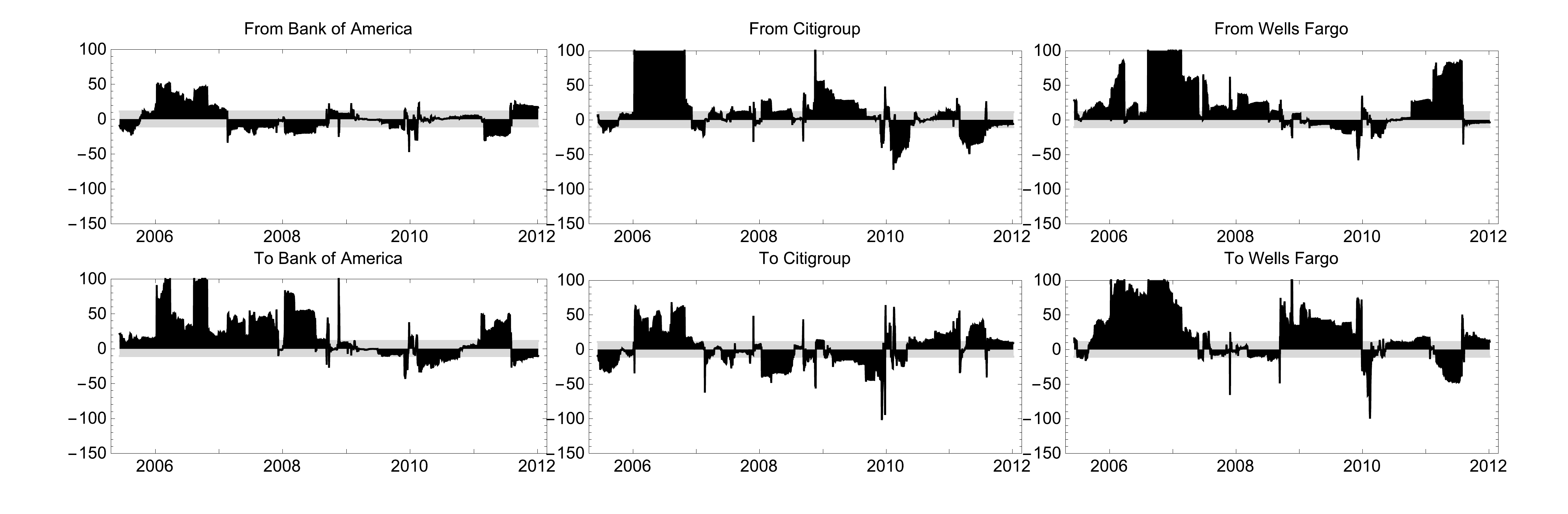} 
   \caption{Directional Spillover Asymmetry Measures $\mathcal{SAM}_{i\leftarrow\bullet}$ and $\mathcal{SAM}_{i\rightarrow\bullet}$ for Financials.} 
   \label{Fig:SAMDIRfin}
\end{figure}

\begin{figure}[h]
   \centering
   \includegraphics[width=5in]{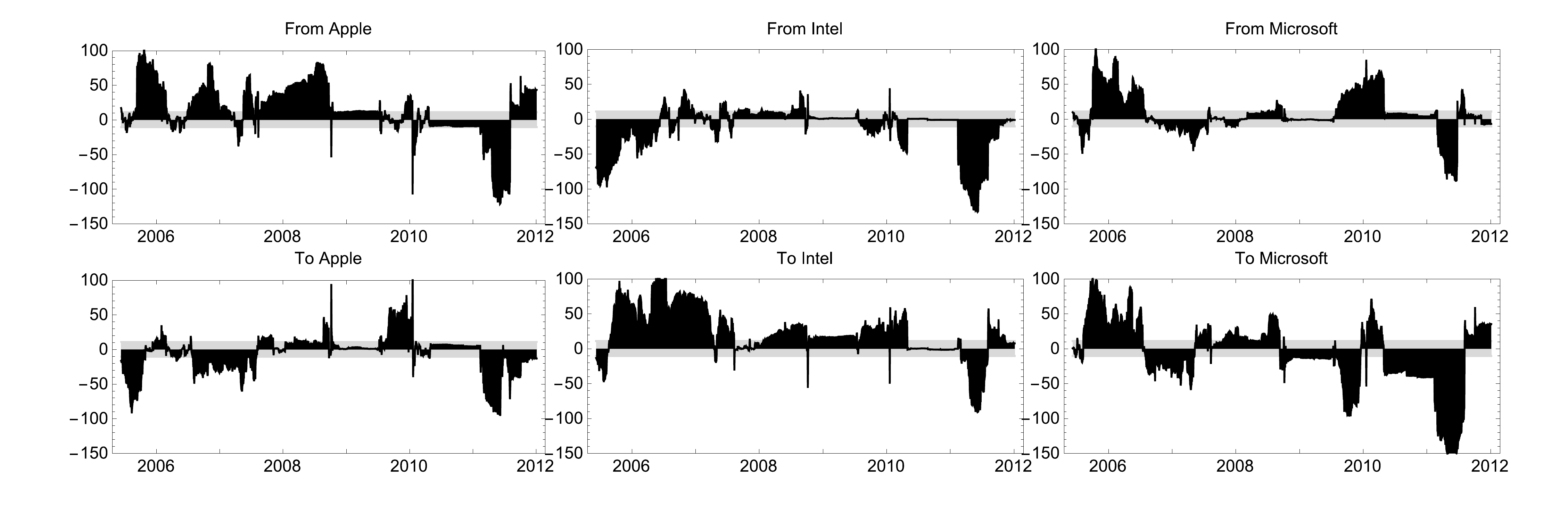} 
   \caption{Directional Spillover Asymmetry Measures $\mathcal{SAM}_{i\leftarrow\bullet}$ and $\mathcal{SAM}_{i\rightarrow\bullet}$ for Information Technology.} 
   \label{Fig:SAMDIRtech}
\end{figure}

\begin{figure}[h]
   \centering
   \includegraphics[width=5in]{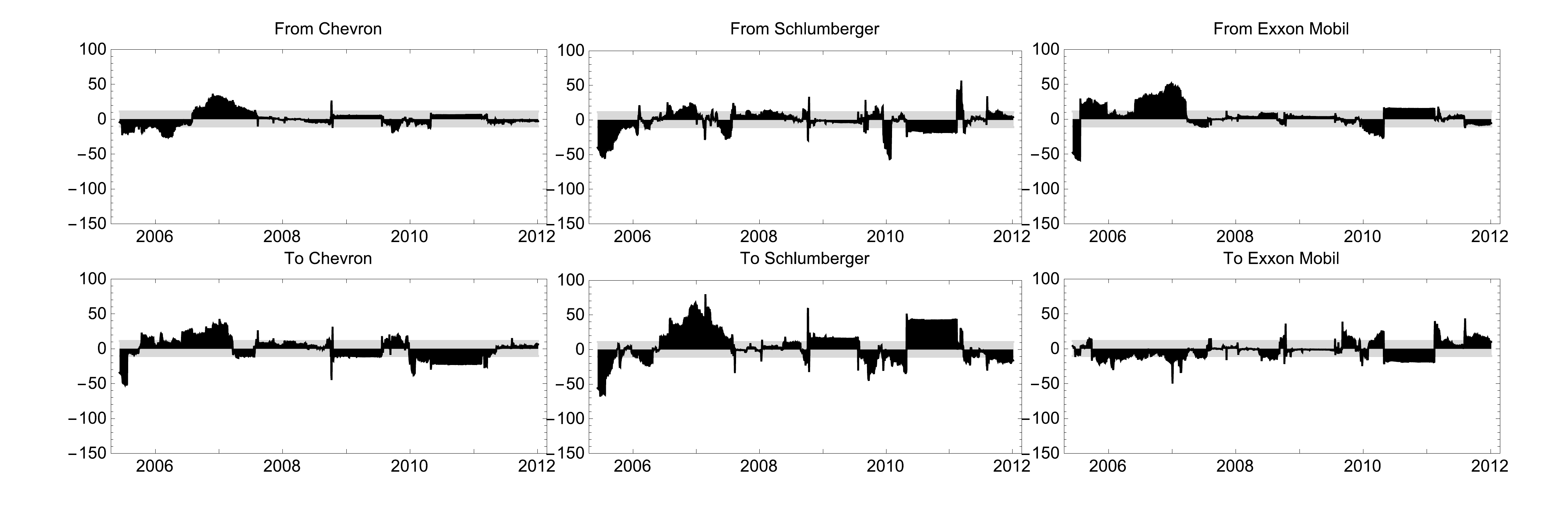} 
   \caption{Directional Spillover Asymmetry Measures $\mathcal{SAM}_{i\leftarrow\bullet}$ and $\mathcal{SAM}_{i\rightarrow\bullet}$ for Energy.} 
   \label{Fig:SAMDIRenergy}
\end{figure}

\begin{figure}[h]
   \centering
   \includegraphics[width=5in]{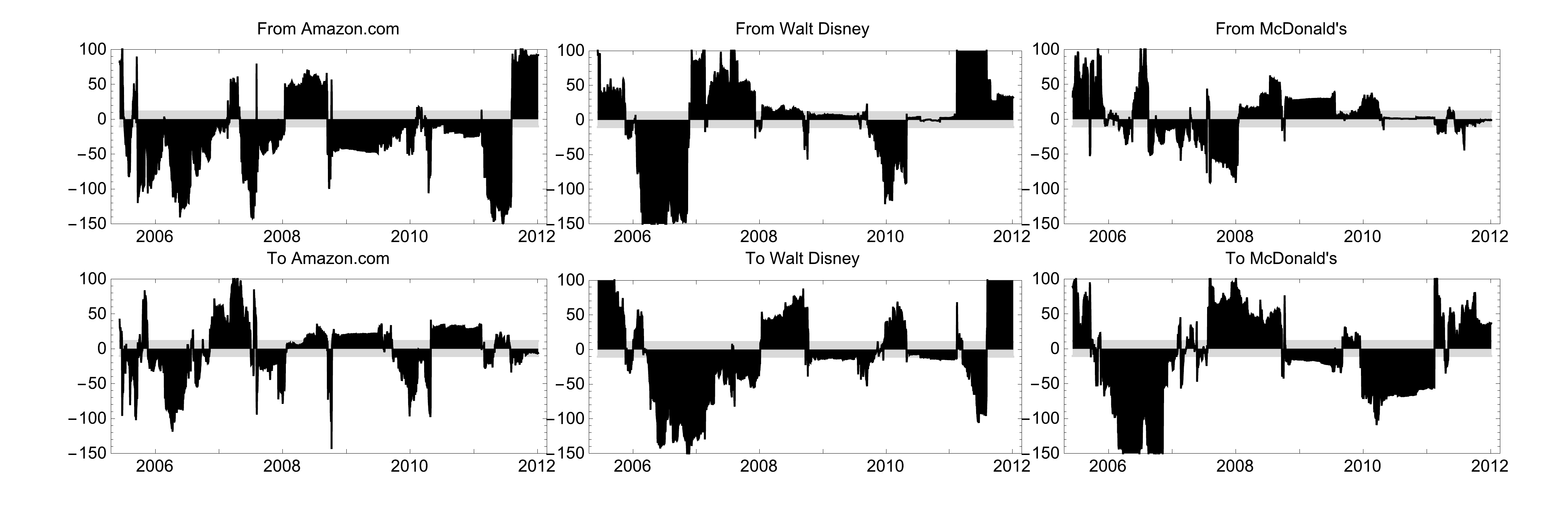} 
   \caption{Directional Spillover Asymmetry Measures $\mathcal{SAM}_{i\leftarrow\bullet}$ and $\mathcal{SAM}_{i\rightarrow\bullet}$ for Consumer Discretionary.} 
   \label{Fig:SAMDIRconscyc}
\end{figure}

\begin{figure}[h]
   \centering
   \includegraphics[width=5in]{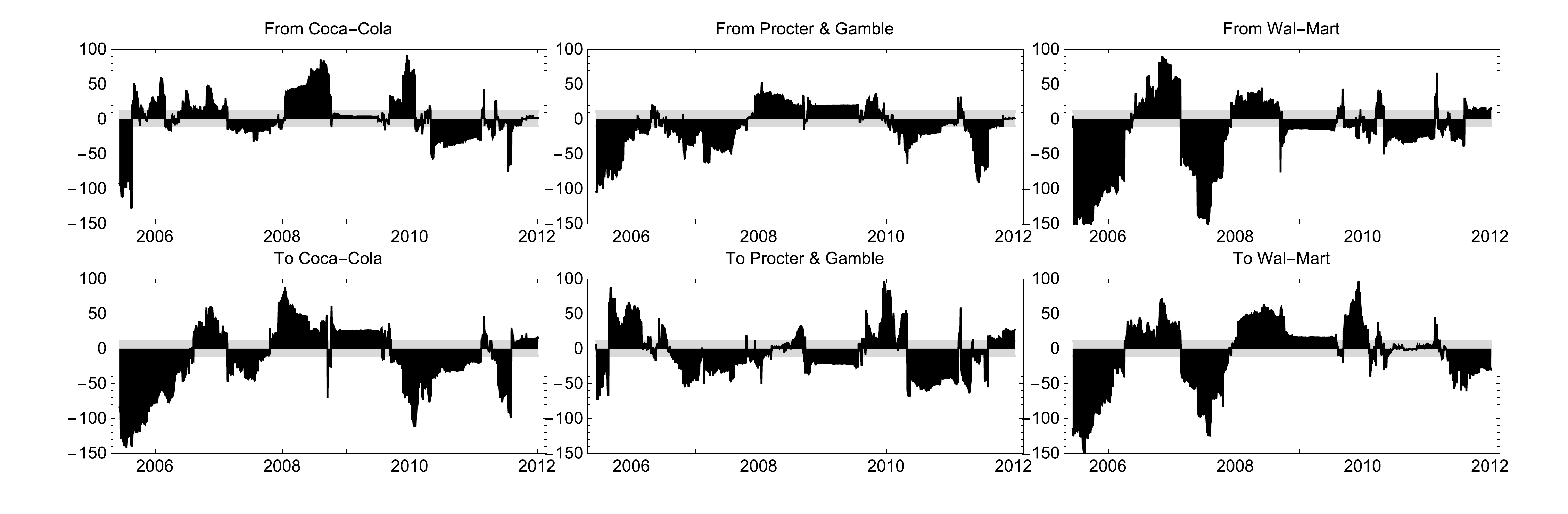} 
   \caption{Directional Spillover Asymmetry Measures $\mathcal{SAM}_{i\leftarrow\bullet}$ and $\mathcal{SAM}_{i\rightarrow\bullet}$ for Consumer Staples.} 
   \label{Fig:SAMDIRconsdef}
\end{figure}

\begin{figure}[h]
   \centering
   \includegraphics[width=5in]{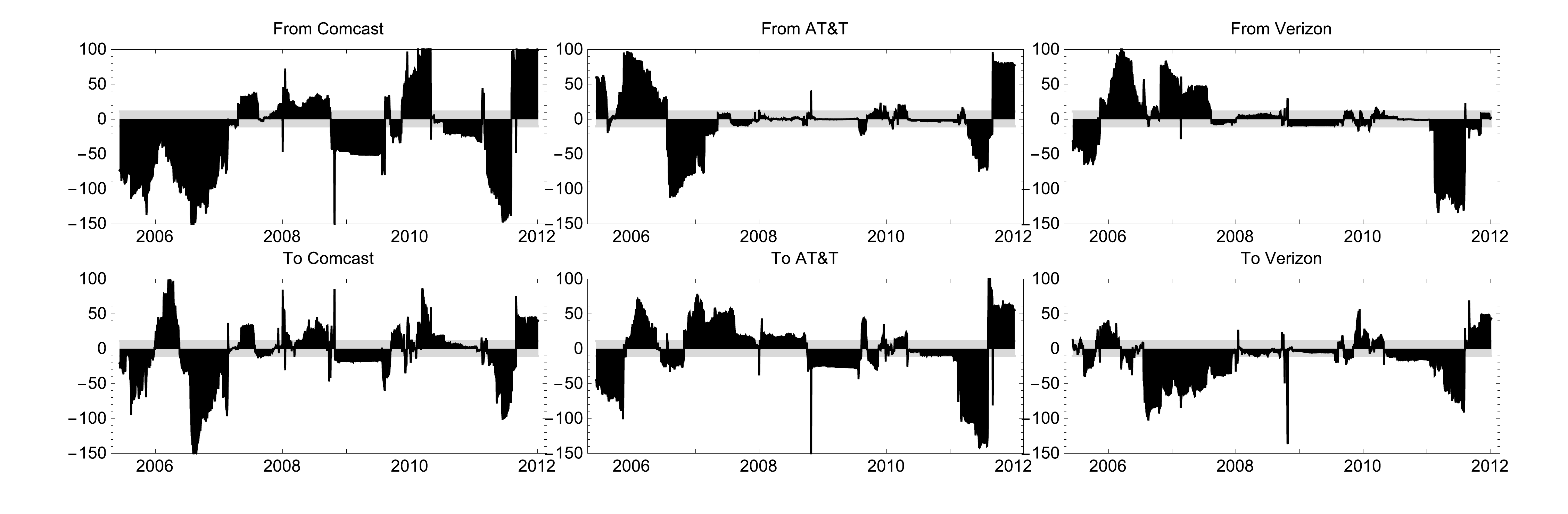} 
   \caption{Directional Spillover Asymmetry Measures $\mathcal{SAM}_{i\leftarrow\bullet}$ and $\mathcal{SAM}_{i\rightarrow\bullet}$ for Telecommunication Services.} 
   \label{Fig:SAMDIRcomm}
\end{figure}

\begin{figure}[h]
   \centering
   \includegraphics[width=5in]{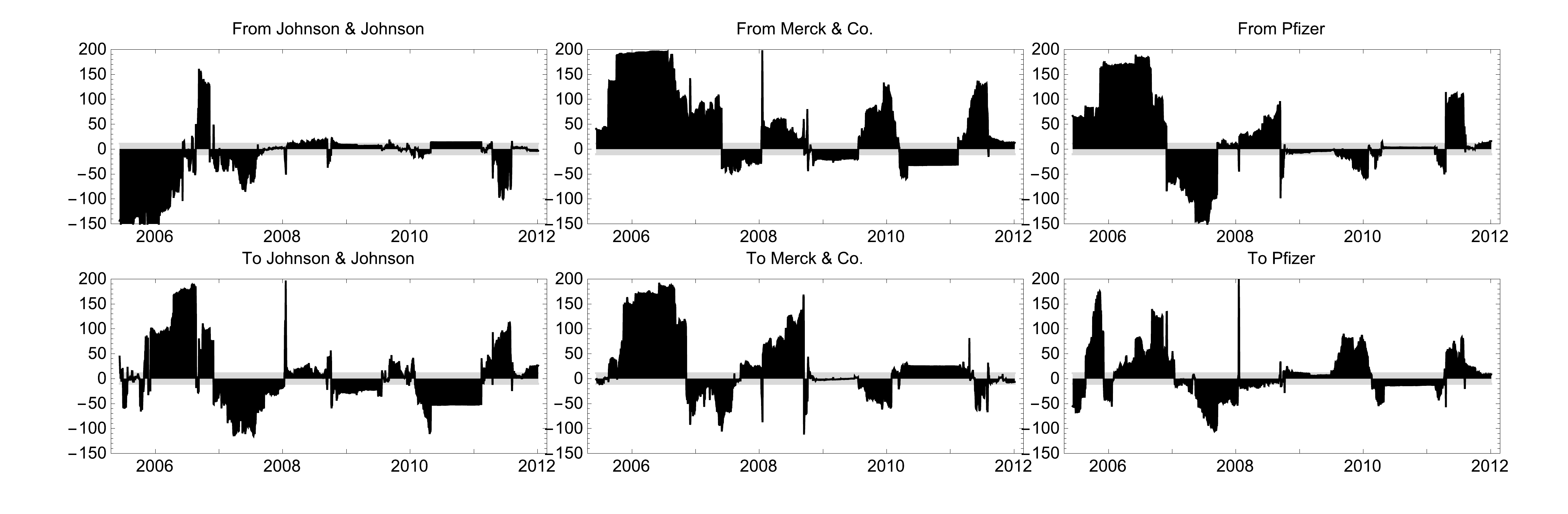} 
   \caption{Directional Spillover Asymmetry Measures $\mathcal{SAM}_{i\leftarrow\bullet}$ and $\mathcal{SAM}_{i\rightarrow\bullet}$ for Health Care.} 
   \label{Fig:SAMDIRhealth}
\end{figure}

To complete our analysis, we study the directional spillovers, which will uncover further information about asymmetries within sectors. Figures \ref{Fig:SAMDIRfin} - \ref{Fig:SAMDIRhealth} show directional spillover asymmetry measures $\mathcal{SAM}_{i\rightarrow\bullet}$ and $\mathcal{SAM}_{i\leftarrow\bullet}$ for the individual sectors.

Whereas $\mathcal{SAM}$ measures the total asymmetry, $\mathcal{SAM}_{i\rightarrow\bullet}$ ($\mathcal{SAM}_{i\leftarrow\bullet}$) identifies the extent to which volatility from (to) the $i$th asset spills over to (from) other assets in the portfolio. If bad volatility from one stock transmits to other stocks more than positive volatility, $\mathcal{SAM}_{i\rightarrow\bullet}$ will be different from zero, and we expect it to be negative. A similar logic applies to $\mathcal{SAM}_{i\leftarrow\bullet}$. Here we again use bootstrapped confidence intervals to test the null hypotheses that $\mathcal{SAM}$ measures are statistically different from zero. We briefly discuss the results for individual sectors in the next paragraphs.

Figure \ref{Fig:SAMDIRfin}, which focuses on the financial sector, is quite informative as it shows that positive spillovers flowing from Bank of America and Citigroup to the rest of the sector ceased right at the peak of the sub-prime mortgage crisis in 2007; in the case of Bank of America the positive spillovers immediately turned into the negative domain.\footnote{\cite{yilmaz2013volatility} reports the direction of the volatility connectedness from U.S. banks toward EU banks in the early stages of 2007, and finds bi-directional evidence after 2007.} Wells Fargo on the other hand did not experience such a sharp drop but kept emanating positive spillovers with a gradual decline until mid-2009. Mainly positive spillovers flew to the Bank of America and Wells Fargo for much of the period, while Citigroup was the recipient of an above-average proportion of negative spillovers. The results are in accord with \cite{billio2012econometric}, who find an increasing level of systemic risk in the finance and insurance industries, with banks playing a much more important role in transmitting shocks than other financial institutions.

Prior to the crisis, excessively large positive spillovers are transmitted from firms in the information technology sector as well. Namely, Apple and Microsoft transmit positive volatility, which is received by Intel (see Figure \ref{Fig:SAMDIRtech}). In a sense Intel is quite an ÒasymmetricÓ stock in that small positive spillovers and large negative spillovers are transmitted from it and an extremely large proportion of the positive spillovers from the rest of the sector is transmitted to it. Over time a frequent load of negative spillovers is transmitted to Microsoft from other technology stocks, with an especially heavy proportion in 2011. In terms of the transmissions from Microsoft the spillovers are relatively symmetric, with a larger proportion of positive spillovers. The pattern is difficult to match with the stock price movements, though.

Directional spillovers show the smallest extent in the energy sector. Just before the crisis, positive spillovers were transmitted from Chevron and ExxonMobil as can be seen in the panels of Figure \ref{Fig:SAMDIRenergy}. This combined outflow of positive spillovers precisely reflects the cumulative perspective shown in Figure \ref{Fig:SAMfin}. ExxonMobil has been continuously receiving disproportionally larger amounts of negative spillovers in the sector, followed by Chevron, which received negative spillovers mainly in 2010.

The crisis period itself is characterized by an alternating pattern of positive and negative spillovers, but the effect of bad volatility is clearly visible and a pessimistic mood dominates most of the period in the consumer discretionary sector. Amazon can serve as an extreme case of evidence (Figure \ref{Fig:SAMDIRconscyc}): negative spillovers are transmitted in clusters from Amazon during almost the whole period, with only a couple of brief exceptions of positive spillovers, and negative spillovers are quite often transmitted to the company as well. Spillovers transmitted from Walt Disney and McDonaldÕs are more balanced overall, but negative spillovers that are transmitted to both companies during 2006 clearly surpass those affecting Amazon. The presented results are fully in line with economic intuition. Consumer cyclical stocks rely heavily on the business cycle and economic conditions. As the state of the economy worsens, disposable incomes shrink and expectations become dismal. The presence of a few positive asymmetries in spillovers thus reflects a fragile economic recovery and shy spending increases during part of the period under research. Clusters of negative spillovers presented in \ref{Fig:SAMDIRconscyc} are massive and further underline the pessimistic mood in this sector.

Asymmetric connectedness is also confirmed in the consumers staples sector, where chiefly negative spillovers are transmitted from Procter \& Gamble and WalMart, while negative spillovers are often unevenly transmitted to Coca-Cola and Procter \& Gamble (Figure \ref{Fig:SAMDIRconsdef}). WalMart as a company is specifically prone to negative spillovers as they are transmitted from and to it during the same time within much of the period under research.\footnote{The massive negative spillovers during 2005 coincide with the decline of WalMart stock in early 2005 and its later stagnation around \$46 into 2007. A further price decline in early Fall 2007 is paralleled in negative spillovers (from and to) during 2007.} As in previous section, the presence of clustering negative spillovers indicates the lack of optimism in the sector that is regularly considered as good hedge in terms of riskiness.

The pattern for the telecommunication services sector is quite irregular (Figure \ref{Fig:SAMDIRcomm}). Negative spillovers are transmitted from Comcast in excess, while AT\&T does not seem to transmit asymmetric spillovers to a great extent for most of the period; those in 2005--2006 and 2011 are exceptions. Verizon transmits positive spillovers well before the financial crisis.\footnote{Negative spillovers transmitted from Verizon in the first half of 2011 match its stagnating stock price development in 2011 after an almost steady increase during 2010.} In terms of the asymmetric spillovers transmitted from AT\&T and Verizon, the sector exhibits either minimal asymmetries in spillovers or enough of the spillovers due to good volatility to be in accord with its defensive nature. However, the economic intuition of the defensive sector or stock does not pair well with Comcast. Clusters of negative spillovers transmitted from the company are rather massive. Based on the overall evidence represented by asymmetries in spillovers, the mood in the sector can be characterized as mildly optimistic, and this is even more so for the period following the initial hit of the crisis.

Finally, we explore the directional spillovers in the health care sector (Figure \ref{Fig:SAMDIRhealth}). Negative volatility transmitted from Johnson \& Johnson counterbalances positive volatility transmitted from Merck and Pfizer prior to the financial crisis. Spillover patterns reflect to some extent the market sentiment that is mirrored in the stock price development.\footnote{The Pfizer stock price followed an inverted U-shape from about \$22 to \$28 and back during the pre-crisis period (2006--2007), and then there was a further gradual drop in price during the financial crisis, until early 2009. MerckÕs stock price offers a similar but more pronounced picture, rising from about \$30 in 2006 at even sharper rate to about \$60 at the end of 2007 with a subsequent fall as the financial crisis fully unfolded. Johnson \& Johnson on other hand did not offer much ground for positive spillovers as prior to the crisis its stock price dropped below \$60 and then oscillated around \$65 with a subsequent drop (from autumn 2008 to early 2009), coinciding with the propagation of the crisis.} During the rest of the time, the health sector exhibits spillover asymmetries in a range similar to the consumer defensive or communication sectors. During the financial crisis substantial negative spillovers dominate the sector with the largest proportion being transmitted by Pfizer. The rest of the researched period is characterized by rather positive spillovers that are transmitted to the sector quite equally, albeit Merck receives positive spillovers in greater proportion. On the sector level, clusters of positive spillovers alter with symmetric performance. Overall, with the exception of the financial crisis, the sector seems to be driven by an optimistic mood. 

To conclude this section, heterogeneity in the sector-specific level turns out to be large, and we cannot see any clear pattern that would hold for all sectors. What we can see is a clear rejection of symmetric connectedness in all sectors, confirming that bad and good volatility are being transmitted with different magnitudes in all sectors.

\section{Conclusion \label{sec:conclus}}
Based on two recent advances in the literature, we outline a simple way to capture volatility spillovers that are due to bad and good volatility. Specifically, we suggest computing the volatility spillover index \citep{diebold2012better} when negative and positive changes in returns are considered separately via realized semivariances \citep{shephard2010measuring}. As a result, we compute volatility spillover indices robust to ordering in VAR that capture asymmetries in volatility spillovers.

We empirically show the versatility of the above set-up by applying it on daily data covering 21 U.S. stocks divided into seven sectors defined in accordance with the Global Industry Specification Standard. We provide ample evidence showing the asymmetric connectedness of markets at the disaggregate sectoral level, which is in contrast to the symmetric volatility transmission mechanism at the aggregate level. The result can be attributed to large sector-level heterogeneity. While there is no clear pattern that would hold for all seven sectors, we are able to reject symmetric connectedness in all of them. Further, we find that bad and good volatility is being transmitted at different magnitudes in all sectors: the consumer, telecommunications, and health sectors exhibit visibly larger asymmetries in spillovers than the financial, information technology, and energy sectors. Finally, we also provide detailed results how asymmetries in spillovers propagate between assets and within sectoral portfolios.

Asymmetries in volatility spillovers have been conclusively detected across the U.S. stock market. While negative asymmetries in spillovers are often of substantial magnitude, they are not strictly dominant. Spillovers due to good volatility materialize quite frequently and their magnitudes are only rarely dwarfed by those due to bad volatility. Hence, in terms of volatility spillovers, market perception is not attuned to negative signals only. Thus, among many detailed inferences, we show that the stock market might be a less dismal place than generally believed.

{\footnotesize{
\setlength{\bibsep}{3pt}
\bibliographystyle{chicago}
\bibliography{spilloversbib}
}}

\addtolength{\topmargin}{-0.5in}
\setlength{\oddsidemargin}{0in}
\setlength{\evensidemargin}{0in}
\setlength{\textheight}{9.4in}

\subsection*{Appendix: Figures}

\newpage

\end{document}